# Enhanced passive thermal stealth properties of VO$_2$ thin films via gradient W doping

**Hyuk Jin Kim,**[a,†] **Young Hwan Choi,**[a,†] **Dongkyu Lee,**[a,b] **In Hak Lee,**[a,c] **Byoung Ki Choi,**[a] **Soo-Hyon Phark,**[d,e] **and Young Jun Chang**[a,b,*]

[a]*Department of Physics, University of Seoul, Seoul 02504, Korea,*

[b]*Department of Smart Cities, University of Seoul, Seoul 02504, Korea,*

[c]*Center for Spintronics, Korea Institute of Science and Technology, Seoul 02792, Korea,*

[d]*Center for Quantum Nanoscience, Institute for Basic Science, Seoul 03760, Korea,*

[e]*Department of Physics, Ewha Womans University, Seoul 03760, Korea.*

**ABSTRACT**

Thermal stealth and camouflage have been intensively studied for blending objects with their surroundings against remote thermal image detection. Adaptive control of infrared emissivity has been explored extensively as a promising way of thermal stealth, but it still requires an additional feedback control. Passive modulation of emissivity, however, has been remained as a great challenge which requires a precise engineering of emissivity over wide temperature range. Here, we report a drastic improvement of passive camouflage thin films capable of concealing thermal objects at near room temperature without any feedback control, which consists of a vanadium dioxide (VO$_2$) layer with gradient tungsten (W) concentration. The gradient W-doping widens the metal-insulator transition width, accomplishing self-adaptive thermal stealth with a smooth change of emissivity. Our simple approach, applicable to other similar thermal camouflage materials for improving their passive cloaking, will find wide applications, such as passive thermal camouflage, urban energy-saving smart windows, and improved infrared sensors.

**Keywords:** Vanadium dioxide ($VO_2$), Thermal stealth, Infrared emission, Metal-insulator transition (MIT), Transmission electron microscopy (TEM)

*Corresponding author.: * yjchang@uos.ac.kr (Y.J. Chang)

† These authors have contributed equally to this work.

# I. INTRODUCTION

Thermal stealth and camouflage concern the hiding of objects, such as military artillery and soldiers, from remote thermal detection devices like infrared (IR) cameras[1–4]. Thermal radiation is proportional to not only surface temperature of an object, but also its thermal radiation emissivity (ε) for a given radiation wavelength. The IR cameras sense thermal radiation in the wavelength range 7.5 to 14 μm, and are capable of detecting objects with elevated surface temperatures in the dark[3]. However, if ε of the object surface is artificially manipulated, the IR detection cannot distinguish a hot object with reduced ε value. Therefore, an ability to control ε would be beneficial for thermal stealth applications.

Electrochromic materials[5–7], optical nanostructures[8,9], plasmonic resonators[10,11], and metal-insulator transition (MIT) materials[12–15], have been reported to enable control of thermal emissivity. In particular, $VO_2$[12,13], $(La,Sr)MnO_3$[16] and rare earth nickelate[15], have been recently suggested to be MIT materials promising for thermochromic switching and urban energy-saving smart window applications[17–19]. MIT is the result of band gap opening at the MIT temperature ($T_{MIT}$), which results in huge changes of ε in IR region (for example, metallic silver (ε = 0.02) and dielectric glass (0.95)) and corresponding changes in surface temperatures measured by IR sensor ($T_{IR}$). Thermal stealth application requires a wide range of active temperature (>30°C) at near the room temperature, however the reported MIT materials have been limited in applications due to their narrow $T_{MIT}$ window (usually, <10°C)[12–15]. To extend the temperature window, active control systems, such as a joule heater, are required[2].

Vanadium dioxide ($VO_2$) is one of the most widely studied MIT materials[12]. In the bulk phase, $VO_2$ undergoes a very sharp MIT at 68°C with MIT width ($W_{MIT}$) of <1°C and a band gap opening of ~0.5 eV. This results in large reversible changes in its IR absorption coefficient and electrical resistance[20], and it is the reason why VO$_2$ films are being considered for thermochromic, thermal stealth[13], smart window[17], and IR-sensing bolometer applications[21]. VO$_2$ thin films exhibit MIT-induced electrical resistivity changes of 2–4 orders of magnitude and W$_{MIT}$ of 5-10°C[22,23]. Variety of oxidation phases (V$_2$O$_3$, V$_2$O$_5$, V$_6$O$_{13}$, and so on) are found to coexist with VO$_2$, and the MIT properties of VO$_2$ show a considerable dependence on the strain and stoichiometry. Meanwhile there have been studies on VO$_2$ thin films doped with tungsten[24–27] or molybdenum[28] to tune their MIT characteristics. Despite enormous efforts to understand the VO$_2$ thin films by means of various growth techniques such as using sol-gel[29], chemical vapor deposition[30], pulsed laser deposition[31,32], molecular beam epitaxy[23], atomic layer deposition[33] and reactive sputtering[34,35], however, a straightforward method of synthesizing VO$_2$ thin films with predetermined T$_{MIT}$ and W$_{MIT}$ values are remained as a challenging task.

In this work, we studied the near-room temperature IR emission properties of W-doped VO$_2$ (W-VO$_2$) films prepared on SiO$_2$, quartz, or mica substrates by vacuum evaporation and subsequent thermal oxidation. In contrast to a narrow emissivity transition temperature window, i.e. W$_{MIT}$ ~5°C, observed in the pure VO$_2$ films, the W-VO$_2$ films showed much enhanced W$_{MIT}$ up to 45°C, which is large enough for practical thermal stealth application. A cross-sectional transmission electron microscopy (TEM) examination of the W-VO$_2$ films revealed that the W concentration is gradually distributed with the distance from the film-substrate interface within W-VO$_2$ layer. We discuss the implications of this observation in terms of the increased W$_{MIT}$ in the W-VO$_2$ films and the large deviation of the T$_{MIT}$ behavior from that in the pure VO$_2$ films as well.

## II. EXPERIMENTAL

We prepared $VO_2$ and W-$VO_2$ films on $SiO_2$/Si, quartz, and mica substrates in two steps. Initially, V or $WO_3$ layers were deposited by electron-beam evaporation at room temperature at a deposition rate of 0.6 Å/s in a high-vacuum chamber (base pressure of $5\times10^{-8}$ torr). For $VO_2$ films, we controlled the thickness of the V layer ($t_V$) from 20 to 100 nm and investigated the effect of $VO_2$ film thickness on IR emission. To form W-$VO_2$ films, a stack of alternating V and $WO_3$ layers (V-W) of 96 nm thick in total were deposited ($t_{V-W}$) by three V (30 nm) and $WO_3$ (2 nm) deposition cycles, which resulted in a tungsten to vanadium ratio of 0.03:1. Both the deposited V and V-W layers were then annealed in the tube vacuum furnace at 400-430°C for 4-5 h (for $VO_2$) or 480-500°C for 1 h (for W-$VO_2$), respectively. During the annealing, an oxygen partial pressure ($PO_2$) of 0.2 torr was maintained with 99.999% $O_2$ gas flow (45 sccm) while pumping the tube furnace with a rotary vacuum pump [35]. The $PO_2$ was chosen after comparing the MIT behavior of $VO_2$ films treated at several different conditions.

Crystallinities of samples were checked by high-resolution x-ray diffraction (HRXRD, Smartlab 9kW, Rigaku) and by cross-sectional TEM imaging with chemical analysis (JEM-ARM200F, JEOL). For TEM, we used a Schottky-type field-emission gun (FEG) operated at 200 keV and prepared cross-sectional specimens by conventional cutting, gluing, polishing, and ion milling (PIPS 691, GATAN) at 5 keV. To study the MIT behaviors of the films, we measured four-probe electrical resistances in the temperature range of 20-100°C by using a closed-cycle refrigerator equipped with a nanovoltmeter (Keithley 2182A), a programmable current source (Keithley 224), and a temperature controller (Lakeshore 330) under a vacuum condition ($<1\times10^{-4}$ torr). An IR camera (FLK-Ti400 (detector resolution 320×240 pixels), FLUKE) was used to measure $T_{IR}$ values of films and bare substrates, while specimens were uniformly heated on a Shamal hot plate (HHP-411, Iuchi Seiei Dou). $T_{IR}$ values of films were determined by comparing them with the temperature of substrates ($T_S$), where $T_{IR}$ values of the hot plate were calibrated with a thermocouple sensor attached to the hot plate for each time. Also, the emissivity was measured by changing the emissivity of the thin films in the IR camera until $T_{IR}$ values of films equals to the hot plate temperature [2].

### III. RESULTS AND DISCUSSIONS

Figs. 1(a) and (b) show HRXRD data of a $VO_2$ film on $SiO_2$/Si and a W-$VO_2$ film on quartz respectively. For both films, various peaks corresponding to different crystalline orientations indicate their polycrystalline character with monoclinic $VO_2$ phase, except negligible fraction of $V_2O_3$ phase. As shown in the inset of Fig. 1(b), the W-$VO_2$ film shows slight shifts and broadening of (002) peaks. Compared to the undoped $VO_2$ film, the W-$VO_2$ film has 0.05% elongation of c-axis. According to the previous studies of thin films, 4 % W-doping is required to induce 0.8% elongation of c-axis and to lower the $T_{MIT}$ by 120 K [25]. The relatively small change of c-axis in our W-$VO_2$ film is probably due to both small doping concentration and possible doping gradient, which also influences broadening of the (002) peak. Figs. 1(c) and (d) show their electrical resistivity curves for increasing (red) and decreasing (blue) temperature sweeps. For the $VO_2$ thin film, the dramatic resistivity changes with large hysteresis indicate the characteristic 1st order MIT behavior. From the sharp MIT temperatures at 71.4°C and 56.0°C during heating and cooling, the hysteresis width, $\Delta T_{MIT} = T_{MIT}$(heating) - $T_{MIT}$(cooling), is calculated to be 15.4°C. Note that $T_{MIT}$ was determined as the temperature corresponding to the inflection point of the resistivity switching curve. The resistivity change of nearly three orders of magnitude observed in the $VO_2$ films was comparable to those reported for both polycrystalline and epitaxial $VO_2$ thin films[22]. On the other hand, the W-$VO_2$ film shows small resistivity change with nearly one order of magnitude, strongly suppressed than that of the $VO_2$ film. According to previous studies of uniformly W-doped $VO_2$ films, W doping gradually reduces the magnitude of resistivity change and also lowers $T_{MIT}$[25,26]. However, the uniformly W-doped ones could not account the small resistivity change in our W-$VO_2$ film. This discrepancy may suggest that the W-$VO_2$ film may have gradual W distribution along the out-of-plane direction, which we will discuss later.

Figure 1. Crystallinity and electrical properties of $VO_2$ and W-$VO_2$ films. HRXRD patterns of (a) $VO_2$ ($t_V$=70 nm) film on $SiO_2$/Si and (b) W-$VO_2$ ($t_{V-W}$=96 nm) film on quartz. Temperature-dependent resistivity of (c) the $VO_2$ and (d) W-$VO_2$ thin films. (Inset of (b) shows comparison of $VO_2$ (002) peaks between $VO_2$ (black) and W-$VO_2$ (red) films.)

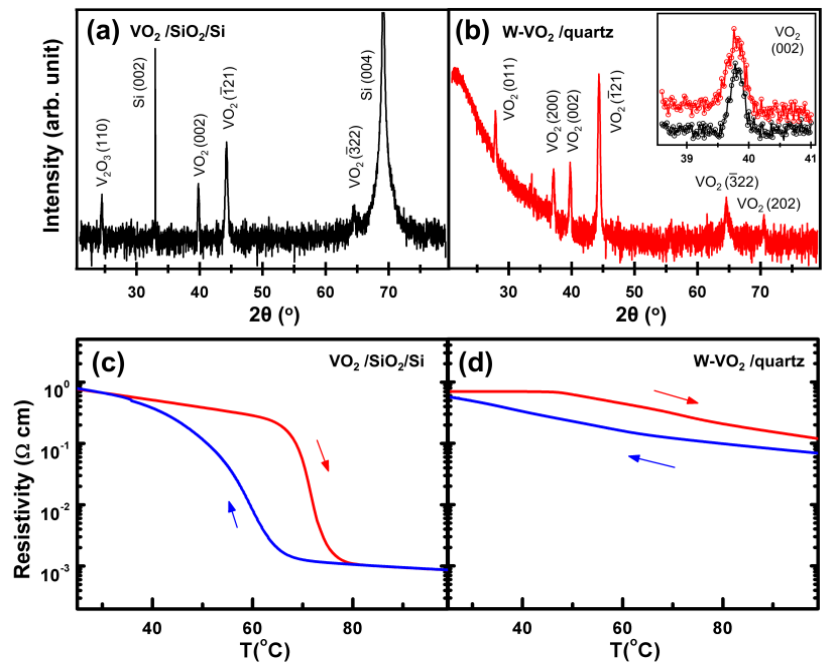

To check the thermal stealth characteristics of the $VO_2$ and W-$VO_2$ films, we captured IR images of the films and substrates on a hot plate (Fig. 2) for an increasing $T_S$ from 35 to 90°C. At a given $T_S$, contrast of the images provided measures of sample $T_{IR}$ values. Figs. 2(a-d) show the IR images of the samples taken at $T_S$ values of (a) 35°C, (b) 60°C, (c) 75°C, and (d) 90°C. At $T_S$ values of 35°C and 60°C (below the $T_{MIT}$=71°C, chosen at the midpoint of the slope for an increasing $T_S$ sweep) of VO$_2$ films, the $T_{IR}$ values of the film were almost the same as that of the SiO$_2$/Si substrate (white arrows in Figs. 2(a) and (b)). At $T_S$ values above the $T_{MIT}$, the $T_{IR}$ values of the films were lower than that of the SiO$_2$/Si substrate (black arrows in Figs. 2(c) and (d)). On the other hand, the W-VO$_2$ film showed a nearly fixed $T_{IR}$ value (sky blue to green color scales, black stars in Fig. 2) regardless of $T_S$ values (35–90°C), with a much wider $T_S$ range than that of VO$_2$ film. Such constant $T_{IR}$ behavior weakens the IR detection of an object, which deserves an improved thermal stealth function. Although previous studies of W doping showed reduced $T_{MIT}$ in VO$_2$ film[25–27], such constant $T_{IR}$ behavior has not been reported for doped VO$_2$ films as far as our best efforts. In addition, all the films exhibit lateral uniformity in color contrast at each temperature, which is related to the lateral homogeneity of film constituents. We note that the color deviation at the perimeters or corners ('α' in Fig. 2(d)) of the samples is because the region is not covered with the films due to clamps during the film growth. Small suppressions in $T_{IR}$ values of the substrates from that of the hot plate (Fig. 2(a): 1.6°C for SiO$_2$ and 1.5°C on quartz) were predominantly due to thermal contact and/or thermal gradient across the substrates.

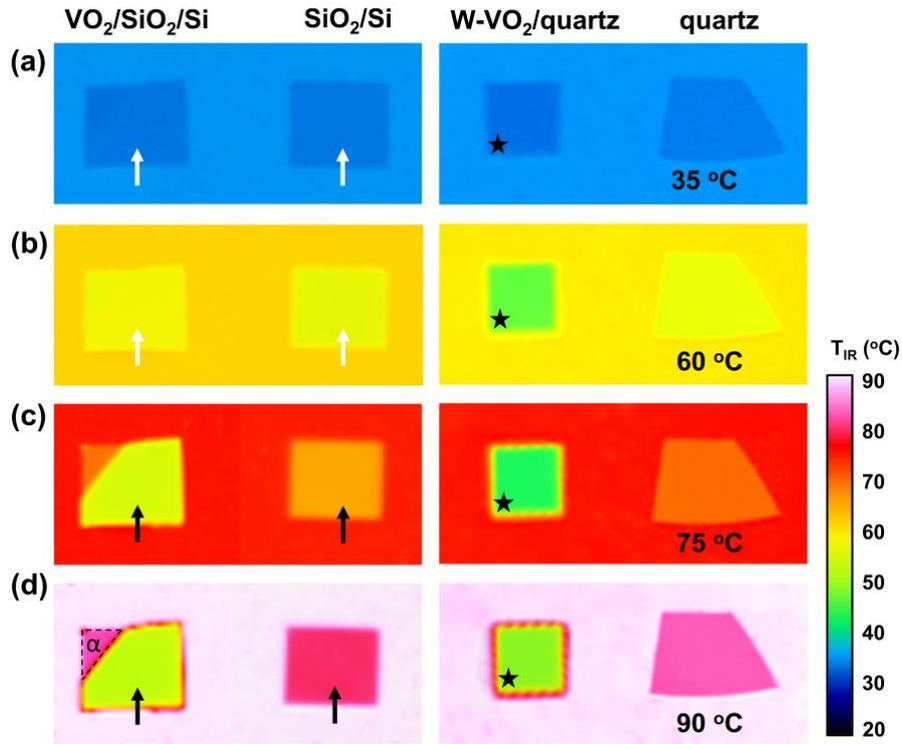

Figure 2. Thermal stealth behavior at different temperatures. (a-d) IR images of the VO$_2$ (left, t$_V$=70 nm) and the W-VO$_2$ film (right, t$_{V-W}$=96 nm) along with their uncoated substrates (SiO$_2$/Si and quartz). The IR images captured at different substrate temperatures (T$_S$); (a) 35, (b) 60, (c) 75, and (d) 90°C.

To obtain quantitative insight of the IR transition behaviors of the VO$_2$ and W-VO$_2$ films, we examined T$_{IR}$ values during heating and cooling sweeps over the T$_S$ range from 25 to 95°C (Fig. 3). VO$_2$/SiO$_2$/Si films showed abrupt T$_{IR}$ transitions at T$_{MIT}$ of 70°C (heating) and 55°C (cooling), which give a ΔT$_{MIT}$ = 15°C (Fig. 3(a)) for different film thicknesses. Furthermore, the temperature difference between T$_S$ and T$_{IR}$ (ΔT$_{IR}$ = T$_S$ - T$_{IR}$) was monotonically enhanced on increasing t$_V$, which lead to a ΔT$_{IR}$ of ~35°C for a VO$_2$ film of t$_V$ = 100 nm at T$_S$ = 80°C (inset of Fig. 3(a)). This observation implies that the increased thickness largely strengthens the magnitude of ΔT$_{IR}$, while change of T$_{MIT}$ is limited within ~5°C.

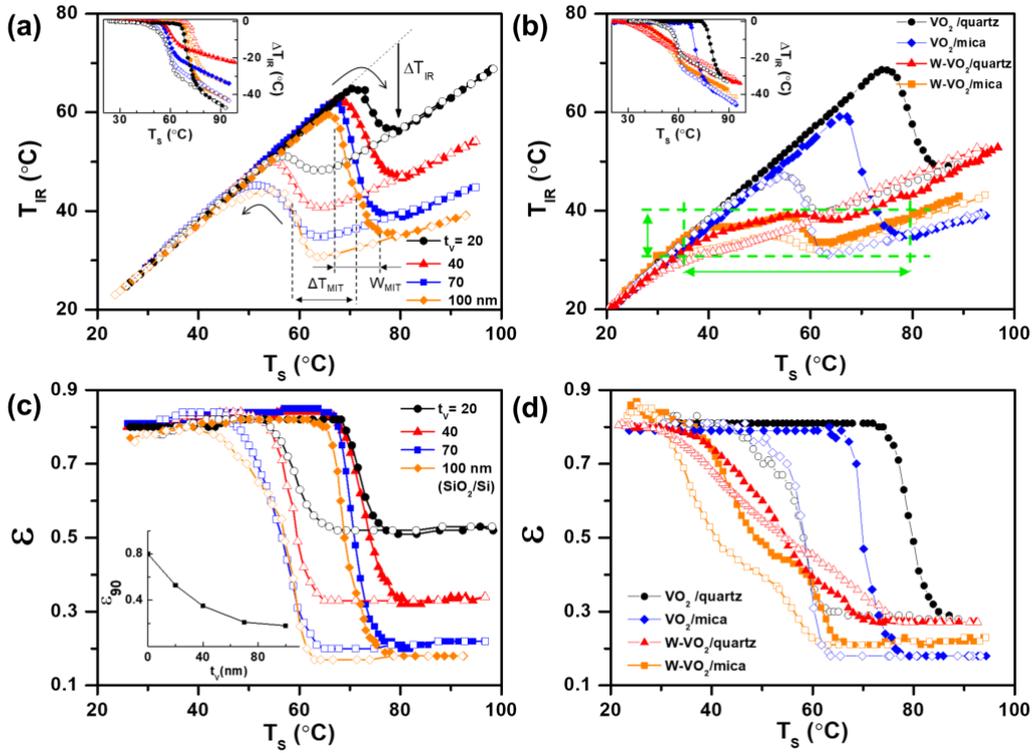

Figure 3. Quantitative analysis of IR detection. Temperature values measured by IR camera ($T_{IR}$) as a function of substrate temperature ($T_S$) for (a) $VO_2$ films grown on $SiO_2$/Si substrates with different thickness ($t_V$ = 20, 40, 70 and 100 nm) and for (b) $VO_2$ and W-$VO_2$ films ($t_{V\text{-}W}$ = 96 nm) grown on quartz and mica substrates. Filled and empty symbols indicate heating and cooling processes, respectively. Insets in (a) and (b) are difference between $T_S$ and $T_{IR}$ of each thin film. Emissivity values for (c) $VO_2$ films on $SiO_2$/Si substrates and for (d) $VO_2$ and W-$VO_2$ films on quartz and mica substrates, extracted from the $T_{IR}$ data in (a) and (b), respectively. Inset in (c) shows thickness dependance of emissivity.

Fig. 3(b) shows $T_{IR}$ versus $T_S$ curves of $VO_2$ ($t_V$ = 100 nm) and W-$VO_2$ films ($t_{V\text{-}W}$ = 96 nm) prepared on quartz and mica. Similar to the films on $SiO_2$/Si, the $VO_2$ films showed sharp $T_{IR}$ transitions with $\Delta T_{MIT}$ of 15–25°C, which were barely observed for the W-$VO_2$ films; the $T_{IR}$ values remained in the range 31 to 40°C (a vertical green arrow) over a wide $T_S$ range of 43°C (a horizontal green arrow). In addition, the broadening of the $W_{MIT}$, led to the collapse of hysteresis; as shown by linear changes in $\Delta T_{IR}$ curves in the inset of Fig. 3(b). It is worth noting that the $\Delta T_{IR}$ values of W-VO$_2$ films became similar to those of VO$_2$ films at $T_S \sim 80°C$, indicative of thermal screening effects similar to that of a VO$_2$ film at a $T_S$ above the $T_{MIT}$. Hence, the W-doping was found to dramatically improve the thermal stealth effect of W-VO$_2$ films by transforming the sharp MIT in the VO$_2$ thin film into the gradual one for a wide temperature range.

Emissivity ($\varepsilon$) of a IR spectrum band can be related a $T_{IR}$-$T_S$ curve via thermal radiance[36], as expressed by: $M(\lambda, T_A) = M_B(\lambda, T_A) \varepsilon(\lambda, T_A) = \frac{2\pi hc^2}{\lambda^5} \frac{1}{e^{hc/\lambda kT_A}-1} \varepsilon(\lambda, T_A)$, where $M$ and $M_B$ are the thermal radiances of material and blackbody, $\lambda$ is wavelength, $h$ is Planck's constant, $c$ is speed of light, $k$ is Boltzmann's constant, and $T_A$ is the absolute temperature. $T_S$-dependent emissivity $\varepsilon(T_S)$, shown in Figs. 3(c) and (d), is extracted from the $T_{IR}$ versus $T_S$ curves in Fig. 3(a) and (b), respectively. $\varepsilon(T_s)$ curves of VO$_2$ films at a given $t_V$ show two saturated values, indicative of two electronic states below and above the MIT. Above the MIT, the $t_V$ dependence of $\varepsilon$ at $T_S = 90°C$ ($\varepsilon_{90}$), shows a saturation behavior approaching ~0.2, indicating an optimal VO$_2$ film thickness ($t_V$) of 100 nm. In Fig. 3(d), we show the $\varepsilon(T_S)$ values of W-VO$_2$ films together with VO$_2$ films, grown on quartz and mica. Here, the $\varepsilon(T_S)$ also saturates at 0.2-0.3, which is similar to that observed for VO$_2$ films above the $T_{MIT}$. However, during MIT in W-VO$_2$ films, the $\varepsilon(T_S)$ curves show a slow decrease over the $T_S$ range of 30-70°C, which explains the monotonic behavior of the $\Delta T_{IR}$ curves in the inset of Fig. 3(b). Furthermore, we note that substrate type has little effect on the features of $\varepsilon(T_s)$ curves.

Fig. 4 shows the microscopic chemical analysis to examine the correlation between the chemical compositions of W-VO$_2$ films and their $T_{IR}$ curves (Fig. 3(b)). Fig. 4(a) shows a cross-sectional TEM image of a W-VO$_2$ film on quartz (t$_{V-W}$ = 96 nm) and its chemical composition profile (Fig. 4(b)). The average film thickness was about 205 nm, which was in-line with thickness estimated from the volume of the VO$_2$ formed from the metallic V film as previously described[37,38]. The chemical composition mappings show laterally uniform distribution of V and W along the in-plane direction. Fig. 4(c) shows averaged concentration profiles for V, W, O, and Si along the dashed arrow shown in Fig. 4(b). Interestingly, W was densely confined towards the film surface while V was distributed rather evenly throughout the whole film thickness. These results clearly show that the doped W were distributed with concentration gradient, where the gradually modulated ratio between W and V is also shown in the inset of Fig. 4(c). Since we initially prepared the alternating layers of V and WO$_3$ at room temperature, the followed thermal oxidation process should play a critical role to form such inhomogeneous concentration distribution. According to the microscopic diffusion model, it should be noted that the oxygen anion diffusion from the surface may promotes a diffusion of cation (probably V) toward the surface to maintain the charge neutrality during the thermal oxidation[39–41]. In the meantime, the oxidation of diffused V forms granular structures, supported with the rough surface morphology in Fig. 4(a). This concentration depth profile supports the notion that the W concentration is indeed modulated and should be related with the exceptionally slow MIT character of our W-VO$_2$ films.

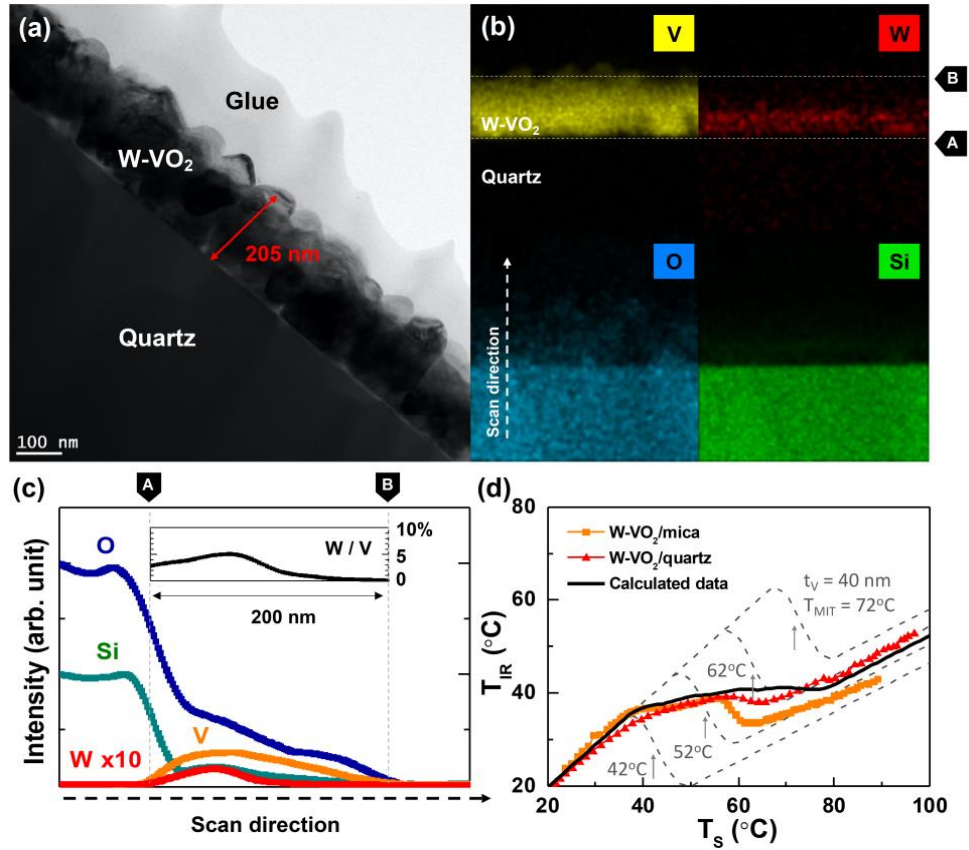

Figure 4. Microscopic analysis of the W-VO$_2$ film. (a) Cross-section TEM image, (b) energy dispersive spectroscopy mapping of four elements (V, W, O, and Si), and (c) elemental depth profiles of a W-VO$_2$ film ($t_V$ = 96 nm) on a quartz substrate. (d) Calculation of $T_{IR}$ versus $T_S$ curve for the W-VO$_2$ film. We duplicated and shifted the temperature measured by IR camera ($T_{IR}$) of $t_V$ = 40 nm sample with 10°C decrement of transition temperature. By averaging the shifted curves, we could fit the experimental curves with the calculated one.

To understand the relation between the modulated concentration profile and the thermal stealth effect, we attempted to calculate the $T_{IR}$-$T_S$ curves of the W-VO$_2$ films with simple assumptions. Previous studies on W-VO$_2$ films have revealed that a homogeneous W doping weakens the MIT characteristics of VO$_2$ films and lowers the $T_{MIT}$ of pure VO$_2$ films in a concentration dependent manner[42,43]. Based on this and the concentration depth profile in Fig. 4(c), we calculated a $T_{IR}$ versus $T_S$ curve for the W-VO$_2$ films with W dopant concentration gradient. First, we choose the $T_{IR}$ versus $T_S$ data of a VO$_2$ film ($t_V$ = 40 nm) as a reference layer. For simplicity, we assume a stack of four W-VO$_2$ layers with four different tungsten dopant concentration and $T_{MIT}$ values (= 72, 62, 52, and 42°C). Those W-VO$_2$ layers are expected to give four $T_{IR}$ versus $T_S$ curves with shifted $T_{MIT}$ values, shown as four dashed gray curves in Fig. 4(d), respectively. By calculating the average of those four different $T_{IR}$ versus $T_S$ curves, we generated a $T_{IR}$ versus $T_S$ curve (a black curve in Fig. 4(d)), which follows the experimental curve for both the W-VO$_2$ films on quartz (a red one) and mica (a pink one). The similarity between the calculated and experimental curves demonstrates that the thermal stealth behavior improved in W-VO$_2$ films might stem from the inhomogeneous W concentration in the films.

Finally, we demonstrated a remarkable improvement of the thermal stealth properties from W-doping-induced attenuation of MIT in VO$_2$ films. On a flexible mica substrate, we also showed the thermal stealth behavior similar to that on the quartz, brightening application of the films highly feasible to curved surfaces. Combined with passive operation without external electrical power, we believe that our findings may provide a low cost and low energy method for a practical application of MIT materials to the military stealth equipment, and passive smart window devices[44,45]. We also showed that such an improvement in the passive thermal stealth characteristics is reproduced also in a stack of VO$_2$ layers with a control of spatial W concentration. Similar improvement of thermal stealth capability has been observed in multilayered thin films[46,47]. In addition, tuning the temperature coefficient of resistance of W-VO$_2$ films may improve the IR sensing performance of the bolometer, a key component for IR-temperature and -image sensors[48].

## IV. CONCLUSION

We fabricated $VO_2$ and $W-VO_2$ thin films by e-beam evaporation with subsequent oxidation and studied the thermal stealth effects of these films by investigating their IR emission properties. The $W-VO_2$ films exhibited the thermal stealth effects at lower temperatures and over a wider range of temperature than the $VO_2$ films, and the microscopic chemical composition analysis revealed that the tungsten dopant was gradually distributed toward the substrate/film interface during the thermal oxidation process. We conclude that the enhanced thermal stealth effects of the $W-VO_2$ films are primarily due to the W gradient doping. Our results suggest a unique solution to overcome the drawbacks of $VO_2$ film stemming from its sharp MIT and large hysteresis characteristics, which may bring the applications of MIT materials to the fields, such as passive thermal camouflage, flexible smart windows, and improved IR sensors, forward.

## Acknowledgements

This work is supported by the National Research Foundation of Korea (NRF) grants funded by the Korea government (NRF-2020R1A2C200373211). This research was supported by Nano·Material Technology Development Program through the National Research Foundation of Korea (NRF) funded by the Ministry of Science, ICT and Future Planning, Korea (2009-0082580). S.-H. Phark acknowledges support from IBS-R027-D1.

**Graphic Abstract**

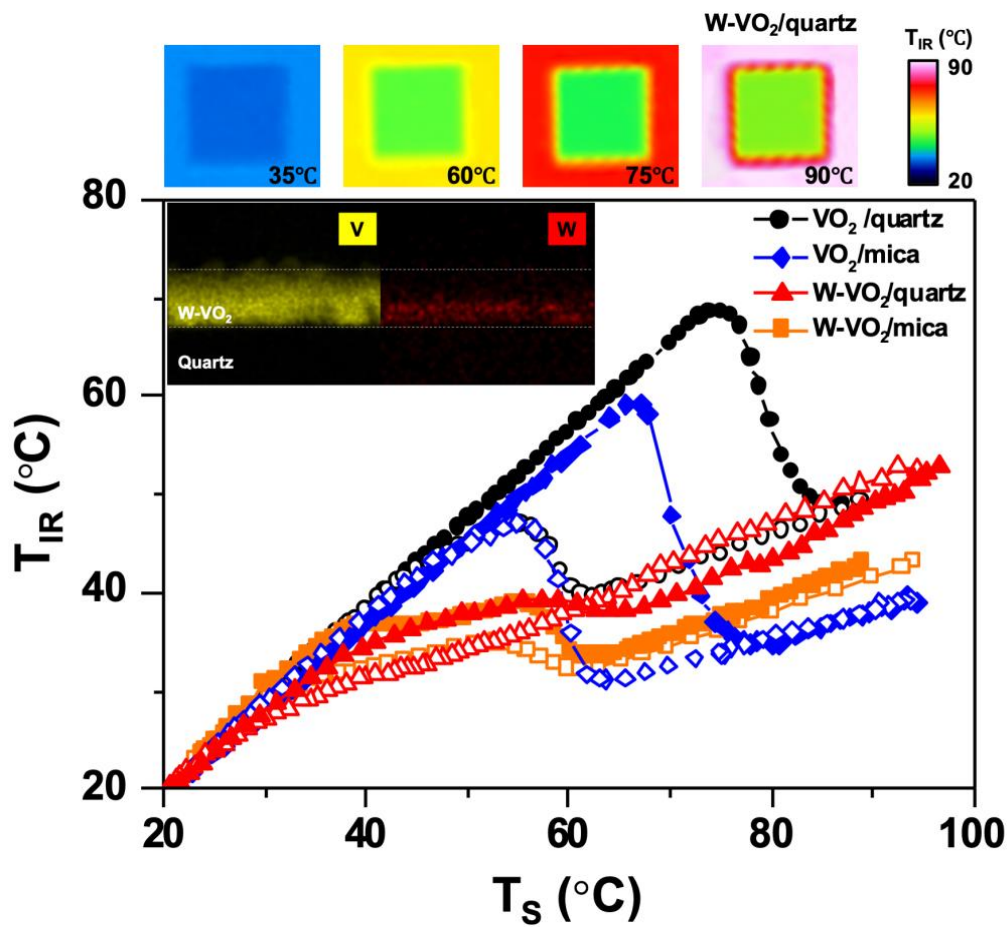

Enhanced passive thermal stealth properties in W-doped VO$_2$ thin films, whose infrared temperature (T$_{IR}$) maintains over wider substrate temperature (T$_S$) range than pure VO$_2$ thin films.

20